\documentclass[aps,showpacs,amsmath,amssymb,twocolumn,nofootinbib]{revtex4-1}

\usepackage{subfigure}
\usepackage{graphicx}
\usepackage{dcolumn}
\usepackage{bm}
\usepackage{color}
\usepackage[colorlinks=true,pdfstartview=FitV,linkcolor=blue,citecolor=blue,urlcolor=blue]{hyperref}
\usepackage[mathlines]{lineno}
\usepackage[dvipsnames]{xcolor}
\usepackage{amsmath}

\everymath{\displaystyle}
\begin{document}

\newcommand{\up}[1]{$^{#1}$}
\newcommand{\down}[1]{$_{#1}$}
\newcommand{\powero}[1]{\mbox{10$^{#1}$}}
\newcommand{\powert}[2]{\mbox{#2$\times$10$^{#1}$}}

\newcommand{\evm}{\mbox{\rm{eV\,$c^{-2}$}}}
\newcommand{\mevm}{\mbox{\rm{MeV\,$c^{-2}$}}}
\newcommand{\gevm}{\mbox{\rm{GeV\,$c^{-2}$}}}
\newcommand{\pgd}{\mbox{g$^{-1}$\,d$^{-1}$}}
\newcommand{\um}{\mbox{$\mu$m}}
\newcommand{\spix}{\mbox{$\sigma_{\rm pix}$}}
\newcommand{\pav}{\mbox{$\langle p \rangle$}}

\newcommand{\sige}{\mbox{$\bar{\sigma_e}$}}
\newcommand{\mass}{\mbox{$m_\chi$}}
\newcommand{\crystal}{\mbox{$f_\textnormal{c}(q,E_e)$}}
\newcommand{\electron}{\mbox{$\rm{e^-}$}}

\newcommand{\gc}[1]{\textcolor{magenta}{[#1]}}
\newcommand{\ms}[1]{\textcolor{blue}{[#1]}}

\newcommand{\beq}{\begin{equation}}
\newcommand{\eeq}{\end{equation}}
\newcommand{\beqs}{\begin{eqnarray}}
\newcommand{\eeqs}{\end{eqnarray}}

\title{XENON1T Anomaly and its Implication for Decaying Warm Dark Matter}

\author{Gongjun Choi,$^{1}$}
\thanks{{\color{blue}gongjun.choi@gmail.com}}

\author{Motoo Suzuki,$^{1}$}
\thanks{{\color{blue}m0t@icrr.u-tokyo.ac.jp}}

\author{Tsutomu T. Yanagida,$^{1,2}$}
\thanks{{\color{blue}tsutomu.tyanagida@ipmu.jp}}

\affiliation{$^{1}$ Tsung-Dao Lee 
Institute, Shanghai Jiao Tong University, Shanghai 200240, China}

\affiliation{$^{2}$ Kavli IPMU (WPI), UTIAS, The University of Tokyo,
5-1-5 Kashiwanoha, Kashiwa, Chiba 277-8583, Japan}
\date{\today}

\begin{abstract}
An excess in the electronic recoil data was observed in the XENON1T detector. One of plausible explanations for the excess is absorption of a vector bosonic particle with the mass of $2-3\,{\rm keV}$. For this, the kinetic mixing $\kappa\!\sim\!10^{-15}$ of the dark photon with the photon is required if the dark photon explains the current DM abundance. We recently proposed a model where the main component DM today is a decaying warm dark matter (WDM) with the life time comparable to the age of the current universe. The WDM decays to a massless fermion and a massive dark photon. This model was originally designed for addressing both the small scale problems which $\Lambda$CDM suffers from and the $H_{0}$ tension. In this letter, we show that the massive dark photon produced by the WDM decay can be identified with the vector boson inducing the anomalous excess in the XENON1T experiment. Depending on a lifetime of the parent decaying WDM, the dark photon could be either the main component or the sub-component of DM population today.
\end{abstract}

\maketitle
\section{Introduction}  
Recently, an incomprehensible excess in the electronic recoil data was observed in the XENON1T experiment~\cite{Aprile:2020tmw}. The energy regime where the excess over the known background appears ranges from 1\,keV to 7\,keV with the events near $2\!-\!3\,{\rm keV}$ particularly prominent. In an effort to interpret the excess as a hint for a new physics beyond the Standard Model (SM), several possible new physics including solar axions, an anomalous neutrino magnetic moment, and bosonic dark matter were discussed and constraints on the relevant physical quantities were reported as well~\cite{Aprile:2020tmw}. (See also the follow-up works about the use of the axion-like particle~\cite{Takahashi:2020bpq} and the dark photon for interpretation~\cite{Alonso-Alvarez:2020cdv} and an possible explanation using an elastic scattering between a particle with the velocity $\sim\mathcal{O}(10^{-2}c$) and the electron~\cite{Kannike:2020agf,Fornal:2020npv} and explanation introducing new interactions between neutrino and electron~\cite{Boehm:2020ltd}.) Among these, in this letter, we pay our special attention to the possibility where the excess is triggered by absorption of a vector bosonic particle~\cite{Pospelov:2008jk} in the XENON1T detector via the dark photon version of photoelectric effect. 

As a resolution to the small scale problems (e.g. core/cusp problem \cite{Moore:1999gc}, missing satellite problem \cite{Moore:1999nt,Kim:2017iwr}, too-big-to-fail problem \cite{Boylan_Kolchin_2011}), the fermionic warm dark matter (WDM) is an interesting possibility. By suppressing the growth of matter fluctuations at scales below its free-streaming length, it may help us understand the observed mass deficit of inner halos in galaxies and galaxy clusters~\cite{Bode:2000gq,Colin:2000dn,Destri:2012yn}. Especially when its mass lies in sub-keV regime and its temperature is low enough, Fermi degeneracy pressure may form to prohibit the core collapse of halos. Aside from the warm nature, this quantum mechanical property is another advantage of the fermionic WDM in regard to the core-cusp problem~\cite{Domcke:2014kla,Randall:2016bqw,DiPaolo:2017geq,Savchenko:2019qnn,Alam:2001sn}.\footnote{The sub-keV mass regime seems to be in severe tension with thermal WDM mass constraint from Lyman-$\alpha$ forest~\cite{Viel:2013apy,Schneider:2018xba}. However, for the case where the WDM has a non-thermal origin, the mass constraint can be relaxed to make the non-thermal sub-keV WDM scenario still viable.} Along this line of reasoning, we recently proposed a consistent model for WDM with a non-thermal origin in Ref.~\cite{Choi:2020tqp}. (See also Ref.~\cite{Choi:2020nan}.) The possibility of decaying WDM was discussed there in which case a fermionic WDM decays to a massless fermion and a massive hidden gauge boson. The model was originally designed for not only the small scale problems but also the Hubble tension.\footnote{Recently it was figured out that decaying DM solution to the Hubble tension is severely constrained by the cosmic microwave background (CMB) power spectrum~\cite{Clark:2020miy}. To enable the late time accelerated evolution of the Hubble expansion rate, the shorter life time and the larger energy transfer from the decaying particle to the radiation decay product are required, which causes inconsistency with the observed CMB power spectrum by increasing the low multipole regime and enhancing oscillations at the high multipole regime. Nonetheless, it might be still probable to avoid this side effects provided a scale dependent spectral index ($n_{s}(k)$) is invoked, which could be still possible depending on an inflation model.} Extending the framework of Ref.~\cite{Choi:2020tqp}, in this letter, we consider a decaying WDM with the lifetime  comparable to or greater than the current age of the Universe. We shall identify the decay product gauge boson with the vector bosonic particle causing the excess in the electronic recoil data recorded by the XENON1T detector. To explain the anomaly, we set the mass of the gauge boson to be $2\!-\!3\,{\rm keV}$ and attribute a significant fraction of the current DM density to the gauge boson. Intriguingly, we found that this set-up motivated by the experimental anomaly provides us with an implication for non-thermal origin of the WDM in the model. This implication is derived from the requirement that effects the decaying WDM has on the CMB anisotropy be minimal for consistency.

\section{A Model}
\label{sec:model}
Having a vector bosonic particle in mind as a source of the anomaly, we introduce an Abelian gauge symmetry $U(1)_{X}$ as an addition to the SM gauge group. The gauge boson $A_{\mu}^{'}$ of $U(1)_{X}$ is to be identified with the vector bosonic particle. For having a fermionic decaying WDM producing $A_{\mu}^{'}$, we introduce chiral fermions charged under $U(1)_{X}$, but neutral to the SM gauge group. Since we are interested in low energy physics (keV regime), we avoid to introduce any vector-like fermion of which a natural mass scale is a UV-cutoff of the theory. Because of the neutrality of dark chiral fermions under the SM gauge group, their $U(1)_{X}$ charges are subject to only two following anomaly conditions
\beq
\sum_{i=1}^{N_{X}}Q_{i}^{3}=0\quad,\quad\sum_{i=1}^{N_{X}}Q_{i}=0\,,
\label{eq:anomalyfree}
\eeq
which are demanding cancellation of $U(1)_{X}^{3}$ anomaly and gravitational $U(1)_{X}\times[{\rm gravity}]^{2}$ anomaly respectively. By referring to \cite{Nakayama:2011dj,Nakayama:2018yvj}, it is realized that the minimum value for $N_{X}$ is five and thus we introduce five chiral fermions with the following exemplary $U(1)_{X}$ charge ($Q_{X}$) assignment 
\beq
\psi_{-9},\quad\psi_{-5},\quad\psi_{-1},\quad\psi_{7}\quad\psi_{8}\,.
\label{eq:U(1)charges}
\eeq
Here the subscripts are denoting $U(1)_{X}$ charges. To make both $A_{\mu}^{'}$ and chiral fermions massive, we introduce two scalars $\Phi_{1}$ and $\Phi_{6}$ with $V_{1}\!>\!V_{6}$ where $V_{Q_{X}}$ is the  vacuum expectation value (VEV) of $\Phi_{Q_{X}}$. Note that even if we introduced two scalars with non-zero VEVs, there is no domain wall problem due to  $V_{1}\!\gg\!V_{6}$ as we see below. We assume the quartic coupling ($\lambda_{Q_{X}}$) of $\Phi_{Q_{X}}$ to be $\mathcal{O}(1)$ so that scalar masses satisfy $m_{Q_{X}}\simeq\sqrt{\lambda_{Q_{X}}}V_{Q_{X}}$. 

Thanks to the charge assignment, the scalars and chiral fermions couple to each other via
\beqs
\mathcal{L}_{{\rm Yuk}}&=&y_{1}\Phi_{1}\psi_{-9}\psi_{8}\, +\,  y_{2}\Phi_{6}\psi_{-5}\psi_{-1}\cr\cr
&+&\,y_{3}\Phi_{6}^{\dagger}\psi_{-1}\psi_{7} \,+\, {\rm h.c.}\,.
\label{eq:yukawa}
\eeqs
Without loss of generality, we can take $y_{i}\,(i=1-3)$ to be real by the field redefinition of chiral fermions. Diagonalizing the mass matrix for $\psi_{-5}$, $\psi_{-1}$ and $\psi_{7}$ yields the following mass eigenstate \beq
\chi\equiv\left(\frac{y_{2}}{\sqrt{y_{2}^{2}+y_{3}^{2}}}\right)\psi_{-5}+\left(\frac{y_{3}}{\sqrt{y_{2}^{2}+y_{3}^{2}}}\right)\psi_{7}\,,
\eeq
which forms a Dirac fermion $\Psi_{{\rm wdm}}=(\psi_{-1},\chi^{*})^{T}$ together with $\psi_{-1}$. Its mass is given by $m_{\rm wdm}\equiv\sqrt{y_{2}^{2}+y_{3}^{2}}\,V_{6}$. We identify $\Psi_{\rm wdm}$ with our WDM candidate hereafter. The remaining  orthogonal direction to $\chi$ becomes a massless Weyl field $\xi$
\beq
\xi\equiv\left(\frac{y_{3}}{\sqrt{y_{2}^{2}+y_{3}^{2}}}\right)\psi_{-5}-\left(\frac{y_{2}}{\sqrt{y_{2}^{2}+y_{3}^{2}}}\right)\psi_{7}\,.
\eeq
This massless $\xi$ would become one of the decay products of $\Psi_{\rm wdm}$. For simplicity, we assume $y_{2}$ and $y_{3}$ to be of a similar order, i.e. $y_{2}\simeq y_{3}\equiv y^{*}$. 

With this basic set-up arranged, we imagine a thermal history of the dark sector which is very similar to the case discussed in our companion paper~\cite{Choi:2020tqp}. To make a long story short, we first introduce a gauge singlet inflaton ($\Phi_{I}$) which couple to scalars in the dark sector at renormalizable level. We assume that the inflaton mainly decays to the SM sector. With the choice of a large enough $V_{1}$ rendering $y_{1}V_{1}$ and $\sqrt{\lambda_{1}}V_{1}$ ($\psi_{-9}$, $\psi_8$ and $\Phi_{1}$ masses) larger than the inflaton mass, at the reheating era the inflaton decay produces only $\phi_{6}$ particles which form a dark thermal bath purely made up of $\phi_{6}$.\footnote{Unless a reheating temperature $T_{\rm RH}$ is close to Planck mass, the quartic coupling induced scattering among $\Phi_{6}$ easily forms a thermal bath. As we shall see later, particles other than $\phi_{6}$ cannot join the dark thermal bath due to the smallness of the gauge coupling $g_{X}$ of $U(1)_{X}$ and Yukawa couplings, $y_{2}$ and $y_{3}$.} Here $\lambda_{1}$ is the quartic coupling of $\Phi_{1}$ and $\phi_{6}$ is the radial direction of $\Phi_{6}$. Since then, the temperature of the dark sector ($T_{\rm DS}$) continues to decrease and $\Phi_{6}$ becomes non-relativistic when $T_{\rm DS}\simeq m_{6}$ is reached. Here $m_{6}$ is the mass of $\Phi_{6}$. With the comoving number density of $\phi_{6}$ preserved, $\phi_{6}$ behaves as a matter until the time comes when $\Gamma(\phi_{6}\!\rightarrow\!{\rm DM\!+\!DM})\simeq H$ holds and $\phi_{6}$ starts to decay to a pair of DMs ($\Psi_{\rm wdm}$) then. With the decay of $\phi_{6}$, the free-streaming of DM candidates $\Psi_{\rm wdm}$ commences. As we shall see, due to $m_{6}>m_{\rm wdm}$, DMs are highly relativistic when it is produced from $\phi_{6}$ and starts free-streaming and thus it is classified as a WDM. We set a life time of $\Psi_{\rm wdm}$ to be comparable to or greater than the age of the present universe so that a significant fraction of $\Psi_{\rm wdm}$ population today is indeed converted into the massless $\xi$ and the $U(1)_{X}$ gauge boson, $A_{\mu}^{'}$. Note that we have made an implicit assumption that scalars in the dark sector have a negligible coupling to the SM Higgs. Also assumed is that kinetic mixing between dark $U(1)$ gauge boson and the SM $U(1)$ is sufficiently suppressed so that dark sector thermal bath has never chance to be in thermal equilibrium with the SM thermal bath. For more details about the thermal history in the dark sector, we refer the readers to Ref.~\cite{Choi:2020tqp}.

\section{XENON1T Anomaly and Dark Photon}
\label{sec:decay}
In the model and the concrete scenario we presented in the previous section, we have the Dirac fermion parent DM, $\Psi_{{\rm DM}}=(\psi_{-1},~\chi^{*})^{T}$, with the mass $m_{\rm wdm}\simeq y^{*}\,V_{6}$. It decays to the massless $\xi$ and the massive $A_{\mu}^{'}$ with mass $m_{A'}=g\,V_{1}$. With the hypothesis that the observed excess in the electronic recoil data in XENON1T stems from absorption of $A_{\mu}^{'}$ via the mechanism analogous to the photoelectric effect, we set $m_{A'}=g\,V_{1}\simeq2-3\,{\rm keV}$. Note that since a large enough value of $V_{1}$ is assumed to integrate out $\psi_{-9}$, $\psi_{8}$ and $\Phi_{1}$, the $U(1)_{X}$ gauge theory comes to be featured by a very weak gauge interaction, i.e. a small $g$ value. 
This structural attribute of the model, in turn, guarantees a large enough life time of $A_{\mu}^{'}$ so that the decay of $A_{\mu}^{'}$ to two massless $\xi$s is suppressed and the produced $A_{\mu}^{'}$ from the decay of $\Psi_{\rm wdm}$ can trigger XENON1T anomaly nowadays. 

\begin{figure}[t]
\centering
\hspace*{-5mm}
\includegraphics[width=0.49\textwidth]{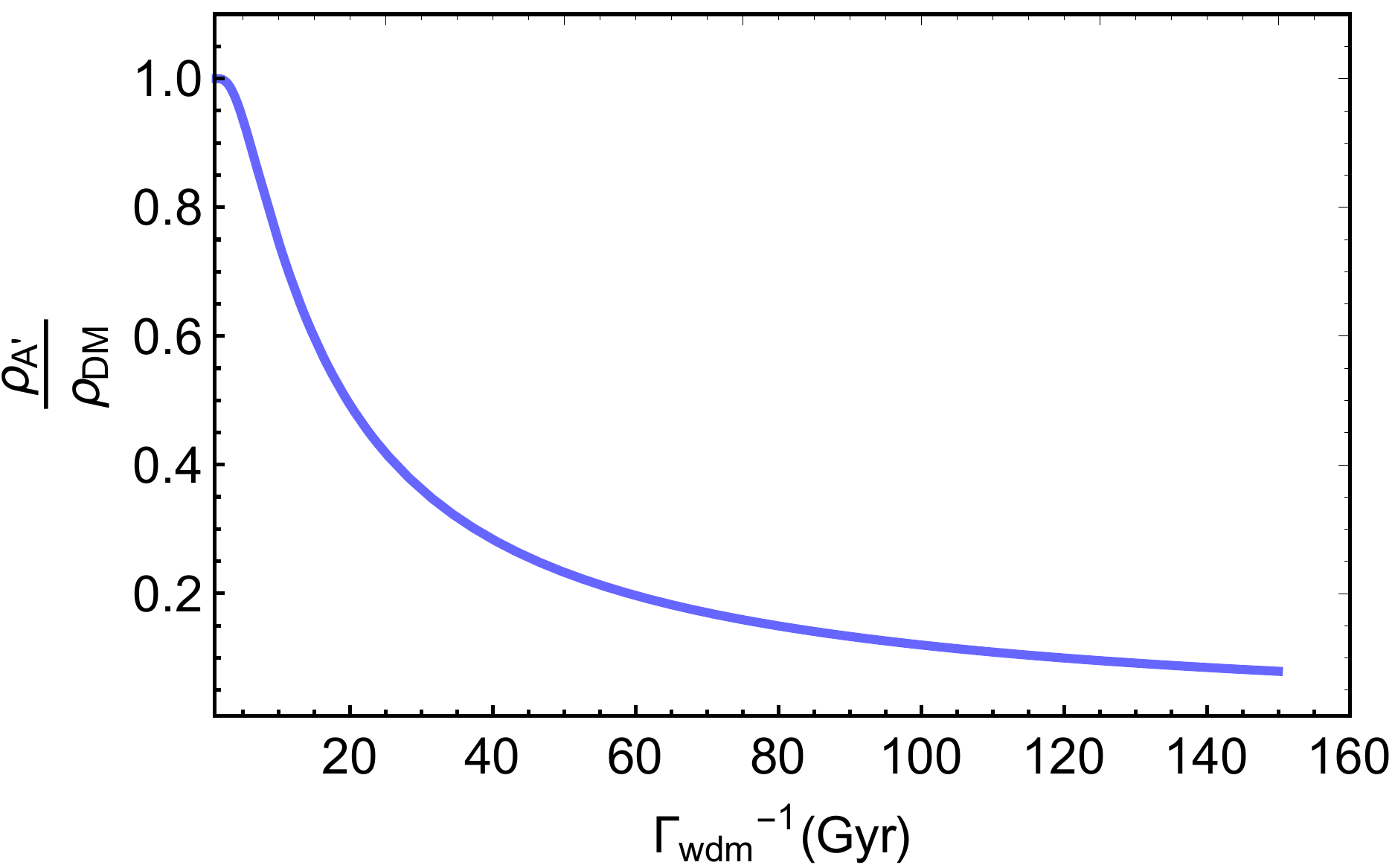}
\caption{The plot of fraction of DM density contributed by $A_{\mu}^{'}$, i.e. $f_{A'}/(1+f_{A'})$ as a function of a life time of decaying WDM. We used $\epsilon\approx0.002(\Gamma_{\rm wdm}/{\rm Gyr}^{-1})^{-0.8}$ which we referred to from Ref.~\cite{Clark:2020miy}.}
\vspace*{-1.5mm}
\label{fig:1}
\end{figure}

We introduce a kinematic parameter $\epsilon$ to quantify a fraction of $m_{\rm wdm}$ transferred to the energy of massless $\xi$. Thus, the massless $\xi$ and the massive $A_{\mu}^{'}$ have the four momenta $p_{\mu}=(\epsilon m_{\rm wdm}, \overrightarrow{p})$ and $p_{\mu}'=((1-\epsilon) m_{\rm wdm},- \overrightarrow{p})$ respectively at the rest of frame of $\Psi_{\rm wdm}$. Now the dispersion relation of $A_{\mu}^{'}$ results in
\beq
\frac{m_{A^{'}}}{m_{{\rm wdm}}}=\sqrt{1-2\epsilon}\,.
\label{eq:massratio}
\eeq
Note that masses of the parent WDM and the decay product $A_{\mu}^{'}$ are almost degenerate in the small $\epsilon$ limit. In terms of $\epsilon$, we can write down $\Psi_{\rm wdm}$'s decay rate, $\Gamma({\Psi_{\rm wdm}\rightarrow A^{'}+\xi})\equiv\Gamma_{\rm wdm}$, as
\beqs
\Gamma_{\rm wdm}&&=\frac{9}{4\pi}\epsilon g_{X}^{2}m_{A'}\left[\left(\frac{m_{\rm wdm}}{m_{A'}}\right)^{3}\!\!-2\frac{m_{A'}}{m_{\rm wdm}}+\frac{m_{\rm wdm}}{m_{A'}}\right]\cr\cr
&&=\frac{9g_{X}^{2}m_{A'}}{2\pi}\times\frac{\epsilon^{2}(3-4\epsilon)}{(1-2\epsilon)^{3/2}}
\,,
\label{eq:GammaDM}
\eeqs
where $g_{X}$ is the gauge coupling of $U(1)_{X}$. By solving the time evolution equation for the energy density of $A_{\mu}^{'}$ and $\Psi_{\rm wdm}$, we find the below expression as the estimate of the energy density ratio $f_{A'}\equiv\rho_{A'}/\rho_{\rm wdm}$ evaluated at today\footnote{For a more exact estimation, see, e.g. Ref.~\cite{Blackadder:2014wpa}}.   
\beq
f_{A'}\simeq e^{\Gamma_{\rm wdm}t_{0}}(1-\epsilon)\left[e^{-\Gamma_{\rm wdm}t_{\rm rec}}-e^{-\Gamma_{\rm wdm}t_{0}}\right]\,,
\label{eq:fraction}
\eeq
where $t_{\rm rec}$ and $t_{0}$ are the age of the universe at recombination and today respectively. From Eq.~(\ref{eq:fraction}), it can be inferred that a set of $(\epsilon,~\Gamma_{\rm wdm})$ determines an abundance of $A_{\mu}^{'}$ today. 

Given XENON1T anomaly, we focus on a value of $\Gamma_{\rm wdm}^{-1}$ which is close to the current age of the universe, i.e. $\sim13.6\,{\rm Gyr}$. With this, we need to avoid inconsistency with the observed CMB anisotropy power spectrum, which could be potentially caused by the decaying dark matter scenario~\cite{Haridasu:2020xaa,Clark:2020miy}. The larger $\Gamma_{\rm wdm}$ and $\epsilon$ we take, the larger amplitude of CMB power spectrum on the low multipole regime and the larger magnitude of oscillations at the high multipole regime would be brought about. Therefore, values of $(\epsilon,~\Gamma_{\rm wdm})$ should be chosen with care so as not to distort CMB power spectrum more than allowed. To this end, we refer to Ref.~\cite{Clark:2020miy} to borrow the expression of the upper-bound $\epsilon\approx0.002(\Gamma_{\rm wdm}/{\rm Gyr}^{-1})^{-0.8}$ ($95\%\,$C.L.). Considering the exemplary range of $150^{-1}\lesssim(\Gamma_{\rm wdm}/{\rm Gyr}^{-1})\lesssim1^{-1}$, we obtain the expected fraction of DM population contributed by $A_{\mu}^{'}$ today as shown in Fig.~\ref{fig:1}. For the exemplary $\Gamma_{\rm wdm}$, we see that $A_{\mu}^{'}$ is responsible for about $\sim\mathcal{O}(10)\%$ of DM energy density today.

Since we are dealing with the spontaneously broken $U(1)_{X}$ ($A_{\mu}^{'}$ is massive), the symmetry of the theory allows for the coupling $U(1)_{X}$ gauge field to the SM through the kinetic mixing with the hypercharge gauge field via
\beq
\mathcal{L}_{\rm kin}=-\frac{\kappa}{2}F^{\mu\nu}F^{'}_{\mu\nu}\,,
\label{eq:kappa}
\eeq
where $F^{'}_{\mu\nu}$ and $F^{\mu\nu}$ are the field strength for $U(1)_{X}$ gauge field and the hypercharge gauge field respectively. Hereafter we use $\kappa$ to denote the kinetic mixing between the dark photon and the SM photon by absorbing $\cos\theta_{w}$ into the $\kappa$ in Eq.~(\ref{eq:kappa}) with $\theta_{w}$ the Weinberg angle. The upper limit for the kinetic mixing between the photon and the dark photon for the mass range $2-3\,{\rm keV}$ of $A_{\mu}^{'}$ reads $\kappa\lesssim10^{-15}$~\cite{Aprile:2020tmw} when the dark photon causing the electronic recoil excess forms the whole DM population today. Now for the case where the lifetime of $\Psi_{\rm wdm}$ is as long as $1\!-\!150\,{\rm Gyr}$, as the decay product of $\Psi_{\rm wdm}$, about $10\!-\!99\%$ of the DM population is attributed to $A_{\mu}^{'}$ as can be seen in Fig.~\ref{fig:1}. Thus the different upper bound on $10^{-15}/\sqrt{0.99}\lesssim\kappa\lesssim10^{-15}/\sqrt{0.1}\simeq(1-3)\times10^{-15}$ applies for our case, depending on a $\Gamma_{\rm wdm}$ value. Note that this is still sufficiently small so that the thermal history of the dark sector we envisioned from the outset remains valid and intact. Note that the choice $150^{-1}\lesssim(\Gamma_{\rm wdm}/{\rm Gyr}^{-1})\lesssim1^{-1}$ was made so that the lifetime of $\Psi_{\rm wdm}$ is not too large to be left with no $A_{\mu}^{'}$ today. 

As for the Hubble tension, the larger $\Gamma_{\rm wdm}$ and $\epsilon$ are favored~\cite{Vattis:2019efj,Clark:2020miy} to make the decaying DM solution to the Hubble tension more viable. However, these parameters cannot be chosen arbitrarily large for consistency with the observed CMB power spectrum~\cite{Clark:2020miy}. Put in another way, the set of values ($\Gamma_{\rm wdm},~\epsilon$)=($1{\rm Gyr^{-1}},~0.002$) being consistent with the observed CMB power spectrum at $95\%$C.L., it cannot solve the Hubble tension since $\epsilon$ is too small to produce $H_{0}$ as large as $74\,{\rm km/sec/Mpc}$. For this point, we speculate that it might be still justified to have $\epsilon$ larger than reported in Ref.~\cite{Clark:2020miy} if a scale dependent spectral index ($n_{s}(k)$) is considered. A reasonable inflation model may be able to achieve a well-designed $n_{s}(k)$ to compensate the problematic change in CMB power spectrum caused by a large $\epsilon$. This may enable a large $\epsilon$ to be still viable to make CMB power spectrum remain consistent with the observed one. If this were to be case, much larger mass of $\Psi_{\rm wdm}$ would be allowed and $H_{0}$ inferred from CMB power spectrum based on the decaying DM scenario~\cite{Vattis:2019efj} might be able to be close to a local measurement of $H_{0}$~\cite{Riess:2016jrr,Riess:2018byc,Bonvin:2016crt,Birrer:2018vtm}

\begin{figure}[t]
\centering
\hspace*{-5mm}
\includegraphics[width=0.46\textwidth]{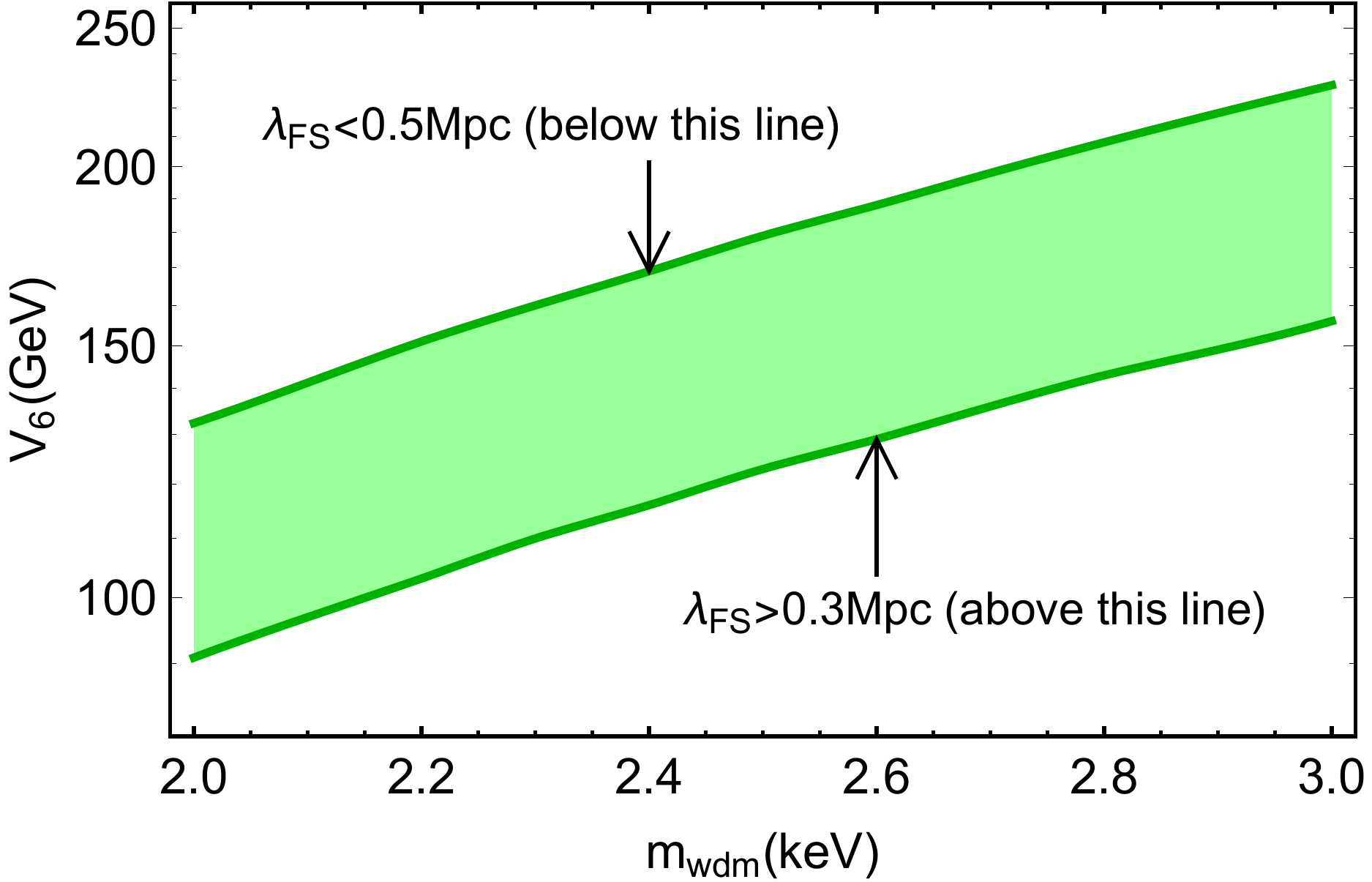}
\caption{The parameter space of ($m_{\rm wdm},V_{6})$ which yields the free-streaming length of interest, i.e. $0.3{\rm Mpc}\!<\!\lambda_{\rm FS}\!<\!0.5{\rm Mpc}$}
\vspace*{-1.5mm}
\label{fig:2}
\end{figure}

\section{XENON1T Anomaly and Decaying WDM}
\label{sec:WDMFS}
Following the discussion of Sec.~\ref{sec:decay}, for $150^{-1}\lesssim(\Gamma_{\rm wdm}/{\rm Gyr}^{-1})\lesssim1^{-1}$, $\epsilon$ parameter is constrained to be at most $\sim10^{-1}$. $A_{\mu}^{'}$ being responsible for the XENON1T anomaly, we can infer the almost equal, but slightly greater mass of the parent WDM $\Psi_{\rm wdm}$ than $m_{A'}\!\simeq\!2-3\,{\rm keV}$ from Eq.~(\ref{eq:massratio}). We note that this mass regime of $\Psi_{\rm wdm}$ is consistent with the mass constraint inferred from the Lyman-$\alpha$ forest observation and redshifted 21cm signals in EDGES observations because $\Psi_{\rm wdm}$ has the non-thermal origin as discussed in Sec.~\ref{sec:model}.\footnote{The current constraint on the mass of the thermal WDM obtained based on Lyman-$\alpha$ forest observation~\cite{Irsic:2017ixq} and redshifted 21cm signals in EDGES observations~\cite{Schneider:2018xba,Lopez-Honorez:2018ipk} are given by $m_{\rm wdm}^{\rm thermal}>5.3\,{\rm keV}$ and $m_{\rm wdm}^{\rm thermal}>6.1\,{\rm keV}$. After mapping these to the mass constraint for a non-thermal WDM originating from the decay of a non-relativistic scalar particle in accordance with Ref.~\cite{Kamada:2019kpe,Choi:2020nan}, we obtain $m_{\rm wdm}\gtrsim 1\,{\rm keV}$.} As such, when $\Psi_{\rm wdm}$ travels the distance of $0.3\,{\rm Mpc}\!<\!\lambda_{\rm FS}\!<\!0.5\,{\rm Mpc}$ since its decoupling, it offers an interesting resolution to missing satellite problem, still achieving consistency with the matter power spectrum at large scales~\cite{Bringmann:2007ft,Cembranos:2005us,Colin:2000dn,Domcke:2014kla}. For the mass regime $2\,{\rm keV}\!\lesssim\! m_{\rm wdm}\!\lesssim\!3\,{\rm keV}$, following the logic in Ref.~\cite{Choi:2020tqp}, we present in Fig.~\ref{fig:2} the space of $V_{6}$ that can lead to the desired free-streaming length. Since $\Psi_{\rm wdm}$ is produced from the decay of the non-relativistic $\phi_{6}$, $\Psi_{\rm wdm}$ starts free-streaming with the initial velocity amounting to $\sim V_{6}/2$. We also checked that $\Delta N_{\rm eff}$ contributed by $\Psi_{\rm wdm}$ during the BBN era is consistent with $\Delta N_{\rm eff}^{\rm BBN}\lesssim0.114$ (95\% C.L.) \cite{Fields:2019pfx}.

\begin{figure}[t]
\centering
\hspace*{-5mm}
\includegraphics[width=0.46\textwidth]{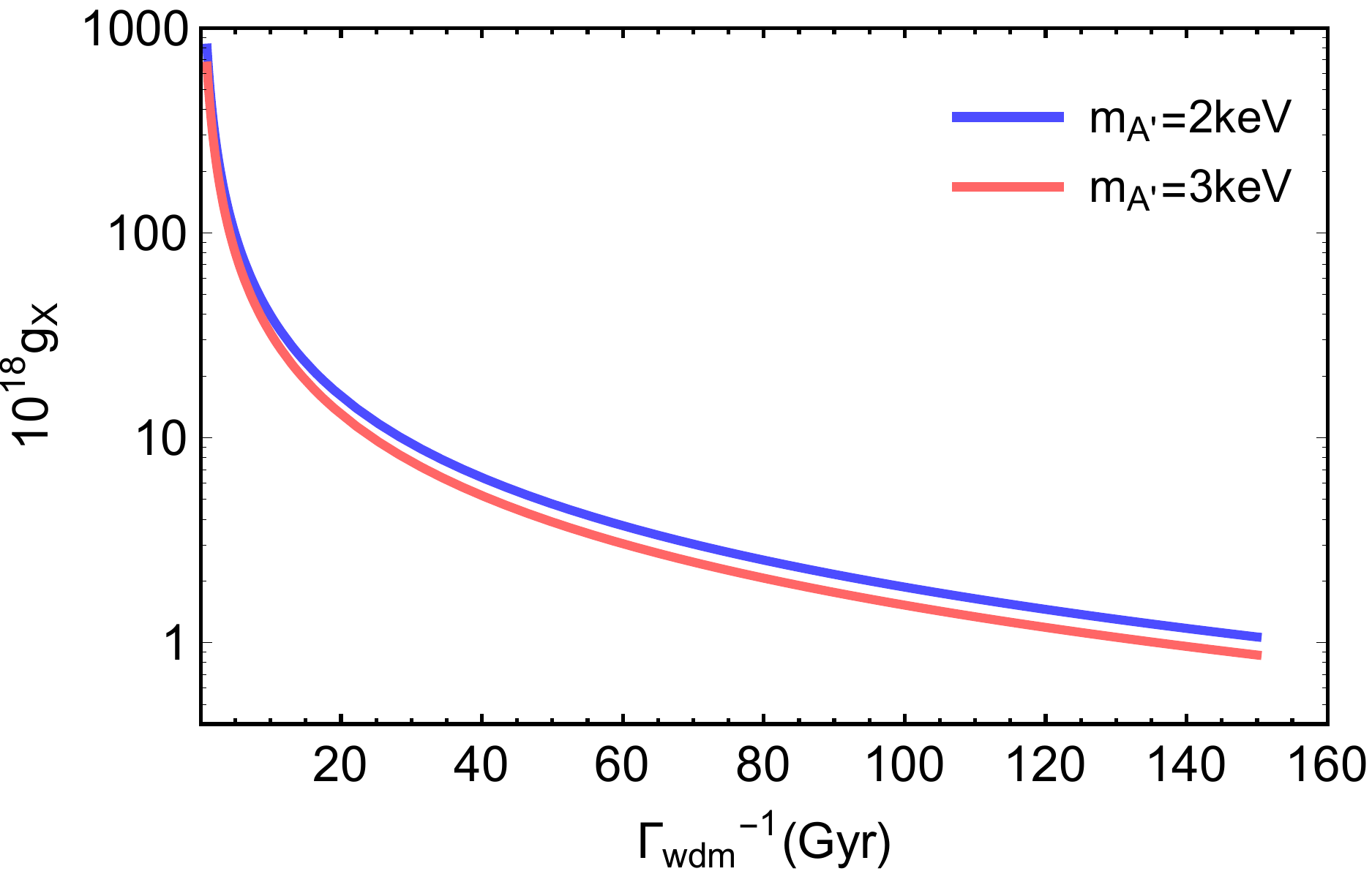}
\caption{The plot of the gauge coupling $g_{X}$ of $U(1)_{X}$ as a function of a life time of decaying WDM.}
\vspace*{-1.5mm}
\label{fig:3}
\end{figure}

Given values of $(\epsilon,~\Gamma_{\rm wdm},~m_{A'})$ of interest, we discuss how the gauge coupling of $g_{X}$ is constrained via Eq.~(\ref{eq:GammaDM}). As discussed in Sec.~\ref{sec:decay}, $\epsilon$ parameter's maximum allowed value is a function of $\Gamma_{\rm wdm}$. Thus, for a fixed $m_{A'}$, Eq.~(\ref{eq:GammaDM}) constrains the gauge coupling $g_{X}$ for the range of $\Gamma_{\rm wdm}$ of our interest. In Fig.~\ref{fig:3}, we show $g_{X}$ as a function of the lifetime of the parent WDM. For $1{\rm Gyr}\lesssim\Gamma_{\rm wdm}^{-1}\lesssim150{\rm Gyr}$, we find that $g_{X}$ is as small as $10^{-18}\!\!-\!\!10^{-15}$. Note that in company with the small Yukawa coupling $y^{*}\!\sim\!\mathcal{O}(10^{-8})$ corresponding to values of $(m_{\rm wdm},V_{6})$ shown in Fig.~\ref{fig:2}, this smallness of the gauge coupling is the underlying reason that guarantees the purity of the dark thermal bath made up of $\phi_{6}$. Also critical is that for $y_{2}$ close to $\sqrt{5/7}y_{3}$, the possible decay process $A_{\mu}^{'}\rightarrow\xi+\xi$ can be sufficiently suppressed thanks to the cancellation in the effective gauge coupling of $\xi$ and the smallness of $g_{X}$.

\section{Discussion}
In this letter, we addressed the problem of explaining a potential source of anomalous excess observed in electronic recoil data taken by XENON1T detector. Attending to the possibility where the excess was caused by absorption of a vector boson, we reduce the problem to pointing out a reasonable candidate of such a vector boson. We previously proposed a decaying fermionic WDM model to deal with the small scale problems and the Hubble tension. Extending the model, we could explain the presence of the desired vector boson by identifying it with the dark photon that naturally arises as one of the decay products of the parent WDM. Hence, the model was shown to be appealing in that it can potentially explain the XENON1T anomaly as well as be invoked for dealing with the small scale problems and the Hubble tension.

In order to have the vector boson today for explaining XENON1T anomaly, we set the lifetime of the decaying WDM to be comparable to or greater than the age of the present universe. The anomaly-inferred mass ($2-3\,{\rm keV}$) of the dark photon, when combined with the consistency with CMB power spectrum, pinned down the similar mass range as the WDM mass, which is consistent with the Lyman-$\alpha$ forest observation and redshifted 21cm signals in EDGES observations due to the non-thermal origin of the parent WDM. As a consequence, the model predicts that a significant fraction of the current DM abundance is attributed to the dark photon. For the free-streaming length range $0.3\,{\rm Mpc}\!<\!\lambda_{\rm FS}\!<\!0.5\,{\rm Mpc}$ of interest and WDM lifetime similar to the age of the universe, couplings ($y^{*},g_{X}$) in the model were shown to be sufficiently small to ensure the thermal history of the dark sector as desired.


\begin{acknowledgments}
T. T. Y. is supported in part by the China Grant for Talent Scientific Start-Up Project and the JSPS Grant-in-Aid for Scientific Research No. 16H02176, No. 17H02878, and No. 19H05810 and by World Premier International Research Center Initiative (WPI Initiative), MEXT, Japan. 

\end{acknowledgments}


\bibliography{main}

\end{document}